\documentclass[twoside,12pt]{article}
\usepackage{amssymb}
\newlength{\dinwidth}
\newlength{\dinmargin}
\newtheorem{theorem}{Theorem}[section]
\newtheorem{prop}[theorem]{Proposition}
\newtheorem{lemma}[theorem]{Lemma}

\newenvironment{proof}{\medskip \noindent 
            {\bf Proof.}}{ \hfill $\square$ \medskip}
\def\idty{{\leavevmode\hbox{\rm 1\kern -.3em I}}}

\def\wnet{{\{\As(W)\}_{W \in\Ws}}}
\def\wrnet{{\{\Rs(W)\}_{W \in\Ws}}}

\def\As{{\cal A}}
\def\Bs{{\cal B}}

\def\Hs{{\cal H}}

\def\Js{{\cal J}}

\def\Ps{{\cal P}}

\def\Rs{{\cal R}}

\def\Ts{{\cal T}}

\def\Ws{{\cal W}}

\def\Zs{{\cal Z}}

\def\Pid{{\Ps_+ ^{\uparrow}}}

\def\idty{{\leavevmode\hbox{\rm 1\kern -.3em I}}}

\def\RR{{\mathbb R}}

\def\IN{{\mathbb N}}
\def\ZZ{{\mathbb Z}}

\begin{document}
\title{An Algebraic Characterization of Vacuum States 
\mbox{in Minkowski Space.\ II.\ Continuity Aspects}}
\author{{Detlev Buchholz}\\
Institut f\"ur Theoretische Physik, Universit\"at G\"ottingen,\\
Bunsenstra\ss e 9, D-37073 G\"ottingen, Germany\\
\vphantom{X}\\
{Martin Florig \ and \ Stephen J.\ Summers } \\
Department of Mathematics, University of Florida,\\
Gainesville FL 32611, USA\\}
\date{\normalsize Dedicated to Eyvind H. Wichmann}
\maketitle
{\abstract An algebraic characterization of vacuum states in 
Minkowski space is given which relies on recently proposed 
conditions of geometric modular action and modular stability for algebras 
of observables associated with wedge--shaped regions. 
In contrast to previous work, continuity properties of these  
algebras are not assumed but derived from their 
inclusion structure. Moreover, a unique continuous unitary representation
of spacetime translations is constructed from these data. 
Thus the dynamics of relativistic quantum systems in Minkowski space 
is encoded in the observables and 
state and requires no prior assumption about any action 
of the spacetime symmetry group upon these quantities.}

\section{Introduction}
Vacuum states in Minkowski space are ordinarily 
characterized by their invariance and 
stability properties with respect to the group of spacetime 
translations. 
This characterization has proven to be a powerful tool,  
both in the structural analysis of relativistic quantum field
theory \cite{StWi,Haag} and in the construction of field theoretic
models \cite{GlJa}. 

In spite of its successes, it seems desirable to reconsider 
this familiar characterization for several reasons. First, there is 
the interesting conceptual problem of whether vacuum states  
can be distinguished in terms of the local observables alone, 
{\it i.e.} without relying on the automorphic action of the spacetime 
symmetry group. An affirmative answer would corroborate the 
view that the full physical information of a theory is  
encoded in the particular ``net structure'' of the 
corresponding observables, {\it i.e.}\ the specific nesting of the 
algebras of observables corresponding to different spacetime 
regions \cite{Haag}. Second, the question is 
of practical relevance in theories on spacetime 
manifolds which do not possess an isometry group as rich as that of
Minkowski space. One is then forced to establish other,
background--independent characterizations of vacuum states,
and a fresh look at the case of Minkowski space theories could  
provide some clues to that effect. Finally, 
in the algebraic approach to the construction of quantum field theories,  
based on universal (free) nets of C$^*$--algebras    
indexed by spacetime regions, the specification of a vacuum state  
amounts to the definition of a theory. One may hope that   
a more intrinsic characterization of vacuum states, relying only 
on the net structure, will also shed new light on these 
constructive problems.  
The conceptual problem of an algebraic characterization of the vacuum state
on Minkowski space, therefore, has received considerable attention 
in recent years, cf.\ the survey articles \cite{Bor4,Sch} and
references quoted therein. 

In the present letter we take up the approach to this problem initiated in 
\cite{BS} and followed up in \cite{Bor2,BDFS}. We shall show 
that the selection criterion for vacuum states proposed in \cite{BS}
can be considerably weakened. Without relying upon any {\em a priori}
assumptions about the presence of symmetries of the net of observables
or on continuity assumptions, we shall show that states 
complying with our weakened criterion 
give rise to continuous unitary representations of the 
translation group which satisfy the relativistic spectrum 
condition. Moreover, these states are ground states for the 
respective dynamics. 
For that portion of the task carried out in \cite{BDFS} 
which we reconsider here, these results also represent a significant 
improvement. 

The mathematical framework and our assumptions  
are specified in the subsequent section, where also a survey
is given of results in \cite{BDFS} which are of relevance here. 
In Sec.\ 3 we construct 
from this input representations of the translations  
and establish the properties indicated above. 
Our letter concludes with remarks on further 
results and open problems. 

\section{The condition of geometric modular action}

In the following, we consider families of algebras 
which are indexed by certain specific wedge--shaped 
regions $W$ of the four--dimensional manifold  
$\RR^4$. These regions can be described with the help
of isotropic vectors $\ell \in \RR^4$ which, in Cartesian coordinates, 
have the form $\ell = (\ell_0, {\vec{\ell}}\,)$ with 
$\ell_0 = |\vec{\ell}\,|$. Given any two such vectors $\ell_\pm$
which are not parallel and a translation $\xi \in \RR^4$, the   
corresponding wedge region $W$ is given by 
$$
W = \{ x \in \RR^4 \mid \pm \, (x - \xi)\cdot\ell_\pm \, > 0 \}.
$$
The set of all these wedges $W$ is denoted by $\Ws$. It is 
stable under translations and Lorentz transformations 
(which map isotropic vectors onto isotropic vectors). We also 
note that each $W \in \Ws$ has a 
complement $W^{\prime} \in \Ws$
given by  
$$
W^{\prime} = \{ x \in \RR^4 \mid \mp \, (x - \xi)\cdot\ell_\pm \, > 0 \}.
$$
It is apparent from these remarks that the manifold $\RR^4$, equipped with the 
distinguished family $\Ws$ of wedges, acquires a natural interpretation as
Minkowski space--time. But we shall not make use of the corresponding 
metric structure in the subsequent investigation. 

Let $\wnet$ be a family of $C^*$-algebras indexed by $\Ws$, each of 
which is a subalgebra of some global unital $C^*$-algebra $\As$. We assume
that this family satisfies the condition of isotony, {\it i.e.}\
$$
\As (W_1) \subset \As (W_2) \ \ \mbox{if} \ \ W_1 \subset W_2,
$$
so it constitutes a net with respect to the partially ordered index set $\Ws$. 
We emphasize that, for the characterization of vacuum states, we do 
not need any further structure on the algebraic side. As a matter 
of fact, the 
wedge algebras may be free algebras without any further relations.

Given the algebra $\As$, the set of positive, linear and normalized
functionals (states) $\omega$ on $\As$ is fixed.
But the states of physical interest form only a 
minute subset of it.  So we have to solve 
here the problem of how to distinguish those states which 
describe the desired vacuum situation. To this end we consider for
each state $\omega$ the corresponding GNS representation 
$(\Hs,\pi,\Omega)$ of $\As$. Within that representation we 
can proceed to the weak closures $\pi(\As(W))''$ of the wedge 
algebras which will be denoted by $\Rs(W)$. 

Our first constraint on the states $\omega$ of interest  
is a condition of Reeh--Schlieder type: for any such state the  
GNS vector $\Omega$ has to be cyclic and separating for all von 
Neumann algebras $\Rs(W)$, $W \in \Ws$. We are then in a position to apply
the results of Tomita--Takesaki theory, see {\em e.g.\/} 
\cite{BratRob}, which yield for each pair $(\Rs(W),\Omega)$
an antiunitary involution $J_W$, called the modular involution, and 
a unitary group $\{\Delta_W^{it}\}_{t \in \RR}$, called the modular
group. The modular objects $J_W$ and $\Delta_W^{it}$ leave $\Omega$ 
invariant and map 
$\Rs(W)$, by their adjoint action, 
onto $\Rs(W)'$ and $\Rs(W)$, respectively, where $\Rs(W)'$
denotes the commutant of $\Rs(W)$ in $\Bs(\Hs)$.  

After these preparations, we can formulate our primary condition 
on the states of interest, the Condition of Geometric Modular 
Action (henceforth, CGMA). It was introduced in the last section of
\cite{BS} and its motivation and significance were explicated at length in
\cite{BDFS}. \\[3mm]
\noindent {\bf Condition of Geometric Modular Action:} \\[1mm]
A state $\omega$ on $\As$ 
fulfills the CGMA if the corresponding net $\wrnet$ 
and vector $\Omega$ satisfy  
\begin{itemize}
\item[(a)] $W \mapsto \Rs(W)$ is an order-preserving bijection,
\item[(b)] if $W_1 \cap W_2 \neq \emptyset$, then 
$\Omega$ is cyclic and separating for $\Rs(W_1) \cap \Rs(W_2)$; 
conversely, if $\Omega$ is cyclic and separating for 
$\Rs(W_1) \cap \Rs(W_2)$, then 
$\overline{W_1} \cap \overline{W_2} \neq \emptyset$, where the 
bar denotes closure, 
\item[(c)] for each $W \in \Ws$, the adjoint action $\mbox{Ad} J_W$ of $J_W$
leaves the set $\wrnet$ invariant, and 
\item[(d)] the group of (anti)automorphisms 
generated by $\mbox{Ad} J_W, \ W \in \Ws$, acts 
transitively on $\wrnet$.  
\end{itemize}
     The first two of these conditions are based on the idea that 
the algebras $\Rs(W)$, $W \in \Ws$, are generated by observables which are 
localized in the respective wedge regions. They establish a   
connection between algebraic properties of the net 
and the lattice structure of the subsets of $\RR^4$, cf.\ the 
discussion  in \cite[Ch.\ III.4.2]{Haag}. The central part of the 
condition is requirement (c), which says that the 
modular conjugations $J_W, W \in \Ws$, generate 
part of the symmetric group on the set $\wrnet$. 
So, in view of the correspondence between algebras and regions, 
we say that they act geometrically.   
Note that no assumptions are made about
the specific form of this action and the nature of 
the resulting group. This very general form of the condition is  
appropriate if one thinks of applications to theories on arbitrary 
spacetime manifolds, 
where $\Ws$ would then be other suitable collections of subregions 
(see \cite{BDFS} for a discussion of the general case). 
The requirement (d) of transitivity is added here for 
simplicity and could be relaxed.   

We shall show in the following analysis that any state $\omega$ 
on $\As$ which satisfies the CGMA for the particular choice of 
regions $\Ws$ made above determines a Minkowski space 
theory which is local and covariant with respect to 
the action of a continuous unitary group of spacetime translations. 
The state itself turns out to be invariant 
with respect to this action, and if it also satisfies the modular stability 
condition, given below, it is a ground (vacuum) state. 
No continuity conditions are
needed for the derivation of this result, in contrast to the 
arguments in \cite{BDFS}. As we shall see, the desired  
continuity properties
are already encoded in the isotony properties of the wedge algebras.   

The first part of our analysis coincides with the discussion in
\cite{BDFS}, which we recall here briefly for the convenience 
of the reader. Assumptions (a) and (c) imply that each $J_W$ 
induces an inclusion-preserving bijection (an automorphism) $\tau_W$ on the 
ordered set $(\Ws,\subset)$ by its adjoint action on the elements of 
$\wrnet$, 
$$
J_W \Rs(W_0) J_W = \Rs(\tau_W(W_0)) \ \ \mbox{for every} \ \
W_0 \in \Ws .  
$$
Since the modular conjugations $J_W$ are involutions, the same is 
true for the corresponding bijections $\tau_W$, $W \in \Ws $.
Moreover, these bijections generate, by composition, a group $\Ts$ of 
automorphisms of $\Ws$ with specific properties resulting from
the CGMA and the modular structure \cite[Lem.\ 2.1]{BDFS}. 
As a consequence of (a) and (b), one has, in
particular, for any $\tau \in \Ts$ and pair of wedges $ W_1,  W_2 
\in \Ws$,
\begin{itemize}
\item[(A)] $W_1 \subset W_2 $ if and only if $\tau(W_1) \subset
  \tau(W_2) $,
\item[(B)] ${W}_1 \cap {W}_2 = \emptyset$ if and only if 
$\tau(W_1) \cap  \tau(W_2) = \emptyset$. 
\end{itemize} 

The concrete form of such automorphisms of the given family $\Ws$ of 
wedges has been determined in \cite[Thm.\ 4.1.15]{BDFS}:\\[1mm]
{\em Any automorphism $\tau$ of $\Ws$ with the properties (A) 
and (B) is induced by an element $\lambda$ of the 
Poincar\'e group $\Ps$ (possibly extended by dilations), i.e.} 
$$
\tau (W) = \{ \lambda x \mid x \in W \} \ \ \mbox{\em for} \ \ W \in \Ws.
$$
This result constitutes a significant generalization of classic results of
Alexandrov \cite{A1,A2} and others \cite{BH,Z}. For the case at hand,
it implies that the group $\Ts$, being generated by involutions, can
be identified with some subgroup $\Ps_\Ts$ of the Poincar\'e
group, {\em i.e.} non--trivial dilations do not occur. 

In a next step, it was shown in \cite{BDFS} 
that subgroups of $\Ps$ which act transitively on $\Ws$ 
have to be large. In fact, one has
\cite[Prop.\ 4.2.9]{BDFS}:\\[1mm]
{\em Any subgroup of $\Ps$ which is generated by a family 
of conjugate involutions and acts transitively 
on $\Ws$ contains the proper orthochronous Poincar\'e group $\Pid$.}\\[1mm]
Because of the
transitivity assumption in the CGMA, the group $\Ps_\Ts$ 
complies with the premises of this result, so it is clear that  
$\Ps_\Ts \supset \Pid$. As a matter of fact, the 
specific properties of $\Ts$ inherited from the modular structure
imply that $\Ps_\Ts = \Ps_+$, the proper Poincar\'e group, 
and the action of the 
elements of $\Ts$ on $\Ws$ is completely fixed.\footnote{If  
the transitivity condition in the CGMA is relaxed, 
then $\Ps_\Ts $ can be one of at most five concrete subgroups of the 
Poincar\'e group \cite{F2}.}   
More concretely, one has \cite[Prop.\ 4.2.10]{BDFS}: \\[1mm]
{\em For any wedge $W \in \Ws$, the corresponding automorphism  
$\tau_W \in \Ts$ is induced by the unique 
involution $\lambda_W \in \Ps_+$ which acts like a 
reflection about the edge of $W$. In particular,
$\tau_W(W) = \lambda_W W = W^{\prime}$, where $W^{\prime}$
is the complement of $W$.}\\[1mm]
So the CGMA fixes the group structure of $\Ts$ 
and the geometric action of its generating elements.
This action is precisely that found by 
Bisognano and Wichmann \cite{BW1,BW2} in their study of the 
modular objects associated with the Minkowski vacuum
and wedge algebras in finite-component quantum field theories satisfying the
Wightman axioms. It is a remarkable fact that 
the much more general class of states complying 
with the CGMA exhibits the same properties. 

Let us return now to  
the modular conjugations $J_W$ which are associated with 
the pairs $(\Rs(W), \Omega)$, $W \in \Ws$. Making use of 
the preceding result, we get  
$$
J_W \Rs(W_0) J_W = \Rs(\tau_W (W_0)) = 
\Rs(\lambda_W W_0), \ \ \mbox{for every} \ \
W_0 \in \Ws , 
$$
hence, in particular, 
$
\Rs(W)' = J_W \Rs(W) J_W = \Rs(\lambda_W W) = \Rs(W') .
$
The latter relation amounts to the following
statement \cite[Prop.\ 4.3.1]{BDFS}:\\[1mm]
{\em The net $\wrnet$ satisfies Haag duality (and thus locality) for
all complementary wedge regions $W, W^{\prime} \in \Ws$.}\\[1mm]
So the CGMA induces commutation properties of the  
net in accord with the causal structure of $\RR^4$ fixed  
by the wedges. It is therefore physically meaningful 
to interpret
the self-adjoint elements of $\Rs(W)$ as observables 
in a Minkowski space theory
which are localized in the wedge regions $W \subset \RR^4$.

Since the modular conjugations associated to $(\Rs(W),\Omega)$ and
$(\Rs(W)^{\prime},\Omega)$ coincide, it follows from Haag duality that
$J_W = J_{W^{\prime}}$ for every $W \in \Ws . $
Hence, as each involution $\lambda_W $  
uniquely determines the pair of wedges $W, {W^{\prime}}$ through
the equation $\lambda_W W = W^\prime$, one can consistently re-label the 
modular conjugations according to $ J(\lambda_W) \equiv J_W$, $W \in \Ws$.
Similarly, if one picks for any other $\lambda \in \Ps_+$ a fixed
decomposition $\lambda = \lambda_{W_1} \cdots \lambda_{W_n}$, one
can define 
$$
J(\lambda) \equiv J_{W_1} \cdots J_{W_n}, \quad \lambda \in \Ps_+.
$$
These (anti)unitary operators generate a group $\Js$ acting upon
$\Hs$. As a matter of fact, one has \cite[Prop.\ 4.3.1]{BDFS}:\\[1mm]
{\em The assignment 
$\lambda \mapsto J(\lambda)$ defines a projective (anti)unitary 
representation of $\Ps_+$ with coefficients in 
a subgroup $\Zs$ contained in the center of $\Js$.
Moreover, the operators $J(\lambda)$, $\lambda \in \Ps_+$, 
leave the vector $\Omega$ invariant 
and act covariantly on the net $\wrnet$, i.e.} 
$$
J(\lambda) \Rs(W) J(\lambda)^{-1} = \Rs(\lambda W) \quad \mbox{\em for} \quad 
W \in \Ws.
$$ 

To summarize, every state $\omega$ on $\As$ which satisfies 
the CGMA determines a local net of wedge algebras 
in Minkowski space on which the proper Poincar\'e
group acts covariantly through some (anti)unitary projective 
representation which leaves the state fixed. This   
result brings us close to our goal,
the characterization of vacuum states in Minkowski space.
What is missing is an argument that the projective representations
obtained in this way 
can be lifted to continuous (true) representations, at least 
for the subgroup of translations. For that is what is needed 
in order to define the energy--momentum content of the states $\omega$
and to address the question under which circumstances    
they are ground states. 

To this end a certain additional continuity condition on the 
net $\wrnet$ was introduced in \cite{BDFS}, and it was 
shown that the desired representations exist in this case. 
In the present paper we drop this technical  
assumption and show that we still obtain  strongly 
continuous unitary representations of the translation subgroup. 
This unexpected result \cite{BS} relies on the geometric 
inclusion structure (isotony) of the wedge algebras. It will 
enable us to define generators of spacetime translations 
and to determine their spectral properties with the help
of a novel modular stability condition proposed in \cite{BDFS}. 

\section{Representations of the Translation Group}

We shall prove now that  
any state $\omega$ on $\As$ which satisfies the CGMA 
for the given set of wedge regions $\Ws$ determines
a strongly continuous unitary representation 
of the translation group $\RR^4 \subset \Ps_+$ which acts 
covariantly upon the net $\wrnet$. The building blocks 
of this representation are products of modular 
conjugations which are associated with shifted wedge regions.

As outlined in the preceding section, the CGMA implies that 
the modular conjugations $J_W$ associated to the pairs  
$(\Rs(W),\Omega)$ induce geometric 
transformations of the net $\wrnet$ which are given by specific 
involutions $\lambda_W \in \Ps_+$, $W \in \Ws$. 
Consequently, the products $J_{W+\xi}\, J_W$, $\xi \in \RR^4$, 
induce the transformations 
$\lambda_{W+\xi} \lambda_W = 
(\xi - \mbox{Ad} \lambda_W \, \xi) \in \RR^4$, {\em i.e.}
pure translations. 

To control the algebraic properties of these products we shall 
make use of the fact that the projective representation 
$J$ of $\Ps_+$, established above, satisfies 
$J(\lambda) \Rs(W) J(\lambda)^{-1} = \Rs(\lambda W)$
and $J(\lambda) \Omega = \Omega$, for $\lambda \in \Ps_+$ and 
$W \in \Ws$. Hence, in view of the uniqueness of the modular 
objects associated with a von Neumann algebra and a faithful 
state, the modular conjugations $J_W$ 
transform covariantly under the adjoint action of the (anti)unitary 
operators $J(\lambda)$, {\em i.e.}
$$
\hspace*{51mm} J(\lambda)J_W J(\lambda)^{-1} = J_{\lambda W}. 
\hspace*{46mm} (\star) 
$$

The essential advance in our argument with respect to the results in 
\cite{BDFS} is the observation that 
the modular conjugations $J_W$ 
enjoy certain continuity properties with respect
to translations of the wedges $W$. 

\begin{lemma} 
Given $W \in \Ws$ and $\xi \in \RR^4$, 
the modular conjugations $J_{W+t\, \xi}$
associated to $(\Rs(W + t \, \xi),\Omega)$, $t \in \RR$, 
are continuous in $t$
in the strong operator topology.
\end{lemma}

\begin{proof} 
Note that, since $W$ is arbitrary, it suffices 
to establish the asserted continuity for $t=0$. 
Consider first the case where 
$\xi$ is such that $W + \xi \subset W$ and, 
to simplify notation, set $J_t \equiv J_{W+t \, \xi}$.
Let 
$\{ t_n \}_{n \in \IN}$ be a decreasing sequence in 
$\RR$ which converges to $0$. Then $\{ W + t_n \, \xi \}_{n \in \IN}$ 
is an 
increasing family of wedges and Lemma 2.6 of \cite{DDLF} implies that the 
sequence $\{ J_{t_n} \}_{n \in \IN}$ converges strongly to the modular 
conjugation $J$ of $(\Rs,\Omega)$, where 
$$
\Rs \equiv \bigvee_n \Rs(W + t_n \, \xi) \, \subset \, \Rs(W) .
$$
In view of the specific geometric action of products of the modular 
conjugations on the wedge algebras, given above, 
and the fact that $W - t_n ( \xi + \mbox{Ad} \lambda_W \, \xi ) = W$, 
one has for any fixed $s > 0$ 
$$
J_0J_{t_n}\,\Rs(W +  s \, \xi) \, J_{t_n}J_0 = \Rs(W + (s - 2t_n)\, \xi) 
\subset \Rs ,
$$
provided $n \in \IN$ is sufficiently large. As $\Rs$ is  weakly 
closed, one can proceed from this inclusion to 
$J_0 J \, \Rs(W +  s \, \xi) \,J J_0 \subset \Rs$ and thence to  
$J_0 J \, \Rs \,J J_0 \subset \Rs$, so that
\begin{displaymath}
\Rs \subset \Rs(W) = J_0 \, \Rs(W)^\prime \, J_0 \subset 
J_0 \, \Rs' \, J_0 
= J_0 J \, \Rs \, J J_0 \subset \Rs .
\end{displaymath}
But this implies $\Rs = \Rs(W)$ and hence $J = J_0$.

Next, let $\{ t_n \}_{n \in \IN}$ be an increasing sequence 
in $\RR$ converging to $0$. Note that $(W + t_n \, \xi)' = W' + t_n \,\xi$, 
so that,
in view of the Haag duality of the net $\wrnet$, one has  
$J_{W + t_n \, \xi} = J_{W' + t_n \, \xi}$. Moreover, 
$\{ W' + t_n \, \xi \}_{n \in \IN}$ is an increasing family of wedges. The 
same argument presented in the first paragraph 
therefore yields the strong convergence 
of $\{ J_{t_n} \}_{n \in \IN}$ to $J_0$.

Finally, let $\{ t_n \}_{n \in \IN}$ be an arbitrary sequence in 
$\RR$ converging to $0$. Since any such sequence contains  
monotone subsequences, for which the strong convergence of the corresponding
modular conjugations has already 
been established, and since the respective limits
coincide, the continuity of the operators $J_t$ at $t=0$ follows
for the special choice of $\xi$.

For arbitrary $\xi$, pick a $\zeta \in \RR^4$ such that 
$W + \zeta \subset W$ and $W + \xi + 2\zeta \subset W$. Then 
$J_{W+t\,  \zeta}$ and $J_{W+t\, (\xi + 2\, \zeta)}$ are continuous in 
$t \in \RR$. According to relation ($\star$)
one has
$$ 
J_{W + t\, \xi} = (J_{W+t\, \zeta} J_{W}) \, 
J_{W+t\, (\xi + 2\, \zeta)} \, (J_{W+t\, \zeta } J_{W})^{-1},
$$
so the strong continuity of $J_{W + t\, \xi}$ follows from the continuity 
properties of the antiunitary involutions appearing on the right--hand side
of this equality.
\end{proof}

With this information we can now proceed as in \cite{BDFS} and show:

\begin{lemma} 
Let $W \in \Ws$ and $\xi \in \RR^4$ be given. The map 
$t \mapsto V(t) \equiv J_{W+t\,\xi} J_W$ is a strongly continuous homomorphism
of $\RR$ into the group of unitary operators on $\Hs$. 
\end{lemma}

\begin{proof} 
Let $J_t \equiv J_{W+t\,\xi}$. As $V(t)$ is product of two such
antiunitary involutions, it is unitary. Moreover, one has 
$V(t) J_0 = J_{t}J_0^2 = J_{t} = J_0^2J_{t} = J_0 V(t)^{-1}$ and 
similarly $V(t) J_{t} = J_{t} V(t)^{-1}$. So 
one obtains, for $n \in \IN$,
$$
V(t)^{2n} J_0 = V(t)^{n} J_0 V(t)^{-n} = J_{2nt} , 
$$
where in the second equality 
relation ($\star$) has been used.
Consequently, one has
$$
V(t)^{2n} = V(t)^{2n} J_0^2 =  J_{2nt} J_0 = V(2nt) . 
$$
Similarly, one finds 
$$
V(t)^{2n+1} = V(t)^{2n} J_{t} J_0 = V(t)^{n} J_{t} V(t)^{-n} J_0 = 
J_{(2n+1)t} J_0 = V((2n+1)t) . 
$$
{}From these relations one sees, in particular, that for $m_1,m_2\in \IN$
and $0 \neq n \in \ZZ$,
$$
V(m_1/n)V(m_2/n) = V(1/n)^{m_1} V(1/n)^{m_2} = V(1/n)^{m_1 + m_2} =
V((m_1+m_2)/n) .
$$
Since $V$ is thus a homomorphism on the subgroup of the rationals and 
is continuous on $\RR$ according to the preceding lemma, it is 
a continuous homomorphism on $\RR$. 
\end{proof}

As the unitary operators $V(t)$ induce the translations 
$t (\xi - \mbox{Ad} \lambda_W \, \xi)$, $t \in \RR$, on the underlying net,  
one obtains with the help of this lemma for every  
one--dimensional subgroup of the translations
a continuous unitary representation. We shall  
show, by using methods developed in \cite{BDFS}, 
that these special representations can 
be put together to a representation of the full 
translation group $\RR^4$. To this end we fix with 
reference to the chosen Cartesian coordinate system the wedges   
$$
W_i \equiv \{ x \in \RR^4 \mid x_i > \vert x_0 \vert \, \}, \quad i = 1,2,3,
$$
and consider, for $\xi \in \RR^4$, the corresponding unitary operators 
$$
U_i(\xi) \equiv J_{W_i + \xi/2} J_{W_i}, \quad  i = 1,2,3. 
$$
These operators induce the translations 
$(\xi - \mbox{Ad} \lambda_{W_i} \, \xi)/2$ on the
net $\wrnet$. Note that if 
$\xi = \xi_0 + \xi_1 + \xi_2 + \xi_3$ is the decomposition of
$\xi \in \RR^4$ into translations along the four
axes of the above coordinate system, there holds in particular 
$(\xi_0 - \mbox{Ad} \lambda_{W_i} \, \xi_0)/2 = \xi_0$ and 
$(\xi_i - \mbox{Ad} \lambda_{W_i} \, \xi_i)/2 = \xi_i$, $i = 1,2,3$.

Let us first consider the restrictions of the 
$U_i(\,\cdot\,)$, $i = 1,2,3$, to the time translations
$\xi_0$. According to the preceding remark and 
Lemma 3.2, these restrictions define 
three continuous unitary representations of the 
one--dimensional group of time translations. 
We show that these representations coincide. 
\begin{lemma}
For all time translations $\xi_0 \in \RR^4$, one has
$$
U_i (\xi_0) = U_j (\xi_0), \quad i,j = 1,2,3.
$$
\end{lemma}

\begin{proof}
The  fact that the rotations in the time--zero hyperplane are induced by 
unitary 
operators in $\Js$ will be employed. If $\rho$ is a rotation by $\pi/2$
about the 1-axis, say, one obtains, first of all, from ($\star$)
the equalities
$$
J(\rho)\,U_1(\xi_0)\,J(\rho)^{-1} = J(\rho)\,J_{W_1 + \xi_0/2}\,
J_{W_1} J(\rho)^{-1} =
J_{\rho\, (W_1 + \xi_0/2)}\,J_{\rho\, W_1 } = U_1(\xi_0) , 
$$
since $\rho\, (W_1 + \xi_0/2)= (W_1 + \xi_0/2)$. 
Next, one notes
that the unitary operators $U_i(\xi_0), \,U_j(\xi_0)$ induce the same
time translation $\xi_0$ on the net. So the differences $U_i(\xi_0) \,
U_j(\xi_0)^{-1}$ map, by their adjoint 
action, each wedge algebra $\Rs(W)$, $W \in \Ws$, onto itself and 
leave the vector $\Omega$ fixed. These differences therefore 
commute with all 
modular conjugations $J_W$, $W \in \Ws$,  
\cite[Cor.\ 2.5.32]{BratRob} and are thus contained in 
the center of $\Js$. In particular, 
$U_1(\xi_0) = Z(\xi_0)\,U_2(\xi_0)$ for some central element 
$Z(\xi_0) \in \Js$. 
Finally, one has
$$
J(\rho) \, U_2 (\xi_0) \, J(\rho)^{-1} = 
J(\rho) \, J_{W_2 + \xi_0/2} \, J_{W_2} \, J(\rho)^{-1} =
J_{\rho \, (W_2 + \xi_0/2)} \, J_{\rho \, {W_2}} = U_3 (\xi_0) , 
$$
since $\rho \,(W_2 + \xi_0/2) = (W_3 + \xi_0/2)$. 
Putting these three facts together, one arrives at the following
relations in $\Js$
\begin{eqnarray*}
Z (\xi_0) \, U_2 (\xi_0) & = & U_1 (\xi_0) = J(\rho) \, U_1 (\xi_0) \, 
J(\rho)^{-1} =  
J(\rho) \, Z(\xi_0)\,U_2 (\xi_0) \, J(\rho)^{-1}  \\ & = & 
Z(\xi_0)\, J(\rho) \, U_2 (\xi_0) \, J(\rho)^{-1} 
= Z(\xi_0) \, U_3 (\xi_0) . 
\end{eqnarray*}
Thus $U_2 (\xi_0) = U_3 (\xi_0)$, and in a similar way one proves 
that $U_1 (\xi_0) = U_3 (\xi_0)$. 
\end{proof}

In view of this result, we can set, for arbitrary 
time translations $\xi_0 \in \RR^4$,
$$
U_0(\xi_0) \equiv U_i(\xi_0), \quad i = 1,2,3.
$$
Next, we consider the operators 
$U_i(\xi_i)$ which, 
according to their geometric action on the net
indicated  above and Lemma 3.2, form a continuous unitary representation
of the one--dimensional subgroups of spatial translations $\xi_i \in \RR^4$,
$i = 1,2,3$. The following result is the final step in our
construction of a unitary representation of the 
full group $\RR^4$ of translations.  
\begin{lemma}
Let $\xi_m$, $m = 0,1,2,3$, be arbitrary translations  
in the four distinguished one--dimensional subgroups of $\RR^4$, 
fixed by the chosen coordinate system. The 
corresponding unitary operators $U_m(\xi_m)$, $m = 0,1,2,3$, commute 
with each other.
\end{lemma}
\begin{proof}
Consider, for example, the operator $U_1(\xi_1)$. It leaves $\Omega$
invariant and satisfies 
$$
U_1(\xi_1) \Rs(W_2 + \zeta) U_1(\xi_1)^{-1} = \Rs(W_2 + \xi_1 + \zeta) 
= \Rs(W_2 + \zeta), \quad \zeta \in \RR^4,
$$
since $W_2 + \xi_1 = W_2$ for all translations $\xi_1$  along the 
1--axis. But this implies that $U_1(\xi_1)$ commutes 
with the modular conjugations $J_{W_2 + \zeta}$ 
and hence with $U_2(\zeta)$, $\zeta \in \RR^4$. 
Thus $U_1(\xi_1)$ commutes in particular with $U_2(\xi_2)$ and since 
$U_0(\xi_0) = U_2(\xi_0)$ according to Lemma 3.3, it also commutes 
with $U_0(\xi_0)$. By the same argument one can establish the 
commutativity of the remaining unitaries. 
\end{proof}

We now define for $\xi = \xi_0 + \xi_1 + \xi_2 + \xi_3 \in \RR^4$
the unitary operators
$$
U(\xi) \equiv U_0(\xi_0) \, U_1(\xi_1) \, U_2(\xi_2) \, U_3(\xi_3). 
$$
According to Lemma 3.2, each of the unitaries appearing on the 
right--hand side defines a   
continuous unitary representation of the corresponding 
one--dimensional subgroup of $\RR^4$. Moreover, 
by Lemma 3.4 these unitaries 
commute with each other. Thus $U$ is 
a continuous unitary representation of the 
group of translations $\RR^4$ and acts geometrically  
correctly on the underlying net. We have 
thus established the following result.
\begin{prop} 
Let $\omega$ be a state on $\As$ which satisfies the CGMA for the particular 
choice of regions $\Ws$ made above.
There exists in the GNS representation $(\pi,\Hs,\Omega)$
induced by $\omega$ a continuous unitary representation 
$U$ of the translations $\RR^4$ which leaves $\Omega$ 
invariant and acts covariantly on the net $\wrnet$, {\it i.e.}
$$
U(\xi) \Rs(W) U(\xi)^{-1} = \Rs(W+\xi) \quad \mbox{for} \quad 
W \in \Ws, \, \xi \in \RR^4.
$$ \label{translations}
\end{prop}

This result is the analogue of 
Lemma 4.3.5 in \cite{BDFS}. 
In view of it, we can turn now to the 
discussion of the energy--momentum spectrum $\mbox{sp}\, U$ 
in the GNS representation induced by $\omega$.
Here again we can rely on the analysis in \cite{BDFS}, whose outcome we recall 
for completeness.   

As $\Omega$ is invariant under the action of $U$, it belongs 
to the point $0$ in the discrete (atomic) part of $\mbox{sp}\, U$. But, 
as was noticed in \cite{BDFS}, the CGMA does not imply that 
$\Omega$ is necessarily a ground state. In fact, there exist
examples fitting into the present framework for which
$\mbox{sp}\, U = \RR^4$. So one has to amend the CGMA
by additional conditions in order to select  
the desired class of vacuum states, for which one 
usually requires that 
$\mbox{sp}\, U \subset \overline{V}_{\! +} \equiv \{\, p \in \RR^4 \mid 
p_0 \geq |\vec{p}| \, \}$ (relativistic spectrum condition). 

Algebraic characterizations of the spectrum condition 
appeared first in 
\cite{Dop} and \cite{Kraus}. More recently, the work of Borchers \cite{Bor1}
(cf.\ also \cite{F1} for simpler proofs) has triggered
renewed interest in this problem \cite{Wi1,BS,Wi2,BrGuLo,GuLo,Wi3}. 
The upshot of these latter investigations is the insight that 
the spectral properties of $U$ are encoded in the 
modular groups $\{\Delta_W^{it}\}_{t \in \RR}$ affiliated 
with the wedge algebras. But in all of these approaches 
the underlying framework was adapted to Minkowski space theories 
and does not seem to allow for a natural generalization to other 
space--times. A  criterion which avoids this problem 
has been proposed in \cite{BDFS}. It also involves 
the modular groups but does not bear explicitly on the specific 
properties of the underlying spacetime manifold.\\[3mm]
{\bf Modular Stability Condition:} \\[1mm]
The modular unitaries are 
contained in the group generated by the modular involutions, {\it i.e.} 
$\Delta_W^{it} \in \Js$, for all $t \in \RR$ and $W \in \Ws$.\\[2mm]
This condition is expressed solely in terms of the algebraically
determined modular objects. Hence, it can also be formulated sensibly for nets
defined over arbitrary space--times, once a collection $\Ws$ of wedge regions 
has been selected. There are already indications from studies 
of nets on de Sitter space--time \cite{BDFS,F2} and also more general 
Robertson--Walker space--times \cite{BS2}
that this condition is indeed relevant to characterize maximally 
symmetric states of particular physical interest.
This is true in spite of the fact that there is no
translation subgroup in the isometry group of these spaces and thus the
standard definition of vacuum state is inapplicable. Furthermore, 
for theories in Anti--de Sitter space the modular stability 
seems to be a characteristic feature of the corresponding vacuua  
\cite{BFS1}. For the case of interest here, Minkowski space, the Modular 
Stability Condition, in conjunction with the CGMA, entails that 
the modular unitaries induce Poincar\'e transformations on the 
wedge algebras (akin to the condition of modular covariance in 
\cite{BrGuLo,GuLo}). As was shown in \cite[Thm.\ 5.1.2]{BDFS}, this 
leads to the following assertion.
\begin{prop} 
Let $\omega$ be a state on $\As$ which satisfies the CGMA, 
with the above choice of wedges $\Ws$,
and the Modular Stability Condition. Then the unitary 
representation $U$ of the spacetime translations whose existence 
has been established in Proposition \ref{translations}
satisfies $\mbox{\rm sp}\, U  \subset \overline{V}_{\! +}$ or 
$\mbox{\rm sp}\, U \subset -\,\overline{V}_{\! +}$. 
\end{prop}
It is a remarkable fact that although neither the CGMA nor
the Modular Stability Condition contains any input about  
the arrow of time (note that the set $\Ws$ is invariant 
under time reflections), every state satisfying these 
two conditions breaks this symmetry. For $\mbox{\rm sp}\, U$ is 
a Lorentz invariant set as a consequence of relation ($\star$) 
and $\mbox{\rm sp}\, U \neq \{0\}$ by part (a) of the
CGMA, so 
the state fixes one of the two cones  $\pm \overline{V}_{\! +}$ 
in the dual of the space--time $\RR^4$. It thereby determines a time direction.
By choosing proper coordinates, we may
therefore assume without loss of generality that 
$\mbox{\rm sp}\, U  \subset \overline{V}_{\! +}$.
With this convention, $U$ is then the  only continuous
unitary representation of the spacetime translations which
acts covariantly on the net and leaves $\Omega$ invariant
\cite[Prop.\ 5.1.3]{BDFS}, 
cf.\ also \cite[Prop.\ 2.4]{BS}. So the apparent ambiguities in our 
construction of spacetime translations have disappeared. 

We have thus arrived at the desired characterization of 
vacuum states in Minkowski space in an algebraic setting 
which is general enough to cover also theories on other spacetime
manifolds. Similar results can be established under slightly 
different conditions. For example, it suffices for the 
proof of the preceding two propositions to assume that 
only even products of the modular conjugations act geometrically on the 
net in the sense of the CGMA \cite{F2}. This result   
is of interest if one wants to include in the 
algebraic setting also non--observable quantities, 
such as Fermi fields, which do not satisfy the condition of 
spacelike commutativity. 
Moreover, as discussed in \cite{BDFS}, there is also a version
of the CGMA based on the modular groups, 
which may be regarded as a generalization of the  
Condition of Modular Covariance, discussed in the literature
\cite{DS,BrGuLo,GuLo,Gu,Da}. We will return to the latter issue elsewhere. 

\section{Further Remarks}

Without any {\em a priori} assumptions of an action of the translation
group upon the net $\wrnet$, we have 
derived from the CGMA and the Modular Stability Condition 
a continuous unitary representation $U$
of the spacetime translations acting covariantly upon 
the net and satisfying the spectrum condition. 
In light of the uniqueness of $U$, we have thus 
determined the dynamics of the system from the given state. 

In \cite{BDFS} it was shown that if the CGMA as well as a certain 
continuity condition of the net $\wrnet$ are satisfied, then there
also exists a strongly continuous unitary representation of 
the full proper Poincar\'e group $\Ps_+$ under which 
$\wrnet$ transforms covariantly. In a forthcoming publication we 
shall extend the arguments given here 
and prove that this strongly continuous representation of $\Ps_+$ can 
also be obtained without assuming any kind of continuity of the 
initial data.

Since we have proven that our conditions are sufficient to
entail that one has a vacuum state on Minkowski space, one may ask
to which extent we have characterized such states. First, the 
work of Bisognano and Wichmann \cite{BW1,BW2} and Thomas and Wichmann
\cite{TW} implies that all of our assumptions are necessary 
if the initial state and algebra 
arise from finite-component quantum fields satisfying the Wightman axioms
and some natural regularity conditions.
It is noteworthy that, from the corresponding nets, this 
underlying field theoretic structure can be recovered in an 
intrinsic manner \cite{FH,SumHPA}. 
Second, it can be deduced from the work of Kuckert \cite{K} that in a vacuum 
state on a net of observable algebras over Minkowski space, if the adjoint 
action of the modular objects maps local algebras onto local algebras, then our
assumptions are again necessary. Therefore, in these physically natural
situations, we have indeed obtained a characterization of vacuum states.

On the other hand, there exist examples of vacuum states
in Minkowski space theories 
\cite{Ar,DS} which do not fit into our setting because 
they are not Lorentz invariant. 
(As already mentioned, our conditions
entail the Lorentz invariance of the  corresponding vacua.)
But these examples are of a rather {\em ad hoc} nature. 
Hence, although from a mathematical point of view
our conditions do not characterize all vacuum states which can 
appear in the algebraic setting of quantum field theory, we 
believe that they distinguish the states of physical interest. 

Let us finally comment on the significance of the choice of the index set $\Ws$
in the CGMA. A detailed discussion of this issue can be found 
in \cite{BDFS}, so we do not need to 
reproduce it here. However, we wish to entice the reader's
interest in this matter with the following remark: If another index set
$\Ws$ of subregions of $\RR^4$ is chosen and a state is found such that 
the CGMA is satisfied, then the group $\Ts$ will, in general, 
be different from the one examined in this paper and 
will, if it can be implemented by point transformations, induce a
different subgroup of the diffeomorphism 
group of $\RR^4$. Interpreting this group as the isometry group of a 
space--time, this suggests that it may be possible to derive geometric
information about a space--time from a net of algebras with a suitable index
set and a state on the net. Indeed, as sketched in the final chapter of
\cite{BDFS}, there emerges the possibility of actually deriving the 
space--time itself from suitable algebraic data satisfying the more general
form of the CGMA presented in \cite{BDFS}.

\end{document}